\documentstyle[12pt]{article}
\newcommand{\be}{\begin{equation}}
\newcommand{\bea}{\begin{eqnarray}}
 
\newcommand{\eea}{\end{eqnarray}}
\newcommand{\ba}{\begin{array}}
\newcommand{\ea}{\end{array}}
\newcommand{\ee}{\end{equation}}

\expandafter\ifx\csname mathbbm\endcsname\relax

\else

\fi
\def\a{\alpha}
\begin{document}
\begin{titlepage}
\hfill
\vbox{
    \halign{#\hfil         \cr
           IPM/P-2002/002 \cr
           hep-th/0202131  \cr
           } 
      }  
\vspace*{20mm}
\begin{center}
{\Large {\bf Supergravity Description of the 
Large $N$ Noncommutative Dipole Field Theories }\\ }

\vspace*{15mm}
\vspace*{1mm}
{Mohsen Alishahiha$^a$ \footnote{Alishah@theory.ipm.ac.ir} 
and  Hossein Yavartanoo$^{a,b}$}
\footnote{Yavar@theory.ipm.ac.ir} \\
\vspace*{1cm}

{\it$^a$ Institute for Studies in Theoretical Physics 
and Mathematics (IPM)\\
P.O. Box 19395-5531, Tehran, Iran \\ \vspace{3mm}
$^b$ Department of Physics, Sharif University of Technology\\
P.O. Box 11365-9161, Tehran, Iran}\\

\vspace*{1cm}
\end{center}

\begin{abstract}
We consider system of D$_p$-branes in the presence of 
a nonzero B field with one leg along brane worldvolume 
and the other transverse to it. We study the corresponding 
supergravity solutions and show that the worldvolume theories 
decouple from gravity for $p\leq 5$. Therefore these 
solutions provide dual description of large $N$ noncommutative 
dipole field theories. We shall only consider those systems which
preserve 8 supercharges in the branes worldvolume. 
We analyze the system of M5-branes and 
NS5-branes in the presence of nonzero C field and  
RR field with one leg along the transverse direction and the others
along the worldvolume of the brane, respectively. 
This could provide a new deformation of (2,0) and little string field 
theories. Finally, we study the Wilson loops using the dual 
gravity descriptions. 
\end{abstract}

\end{titlepage}

\section{Introduction}
It was believed for a long time that the four dimensional 
large $N$ gauge theory have a string theory description
\cite{THO}, though the strings live in more than four 
dimensions \cite{POL}. The well-known example of this 
correspondence is ${\cal N}=4$ $SU(N)$ SYM 
theory in four dimensions whose string theory dual is 
ten dimensional superstring theory known as type II B 
\cite{{MAL}, {GKP}, {WIT}}. 
By now we know a large class of examples in the context of
AdS/CFT correspondence (see \cite{MAGOO} for a review) 
which relates field theories without gravity to superstring 
(string) theories on certain curved background. This correspondence 
naturally arises when considering 
D$_p$-branes in a limit where the worldvolume field theory 
decouples from the bulk gravity. 

This correspondence has been generalized for 
the cases where the field theory is non-local, 
{\it e.g.} for non-commutative field theories. 
From string theory side, these theories can be obtained 
on the D-brane worldvolume in the presence of a non-zero 
B field \cite{DH}. With $N$ coinciding D$_p$-branes in 
the presence of a non-zero B field the worldvolume theory is 
deformed to a $U(N)$ noncommutative SYM theory \cite{SW}.

As in the case with $B=0$, there exists a limit where the 
bulk gravity decouples from the worldvolume noncommutative 
field theory \cite{{SW},{CDS}} and a correspondence between
string theory on curved background with B field and 
noncommutative field theory is expected. Actually this issue
has been investigated during last three years in several papers
including \cite{ALL}-\cite{ALL6}.

Turning on a B field on the D-brane worldvolume can be viewed,
via AdS/CFT correspondence, as a perturbation of the 
worldvolume field theory by an operator of dimension 6. 
For example in the D3-brane case, from the four dimensional 
superconformal Yang-Mills
theory point of view the bosonic part of the dimension 6 operator
is given by \cite{{DT},{FLZ}}
\be
{\cal O}_{\mu\nu}={1\over 2g^2_{\rm YM}}{\rm Tr}\left(
F_{\mu \delta}F^{\delta \rho}F_{\rho \nu}
-F_{\mu\nu}F^{\rho\delta}F_{\rho \delta}+2
F_{\mu\rho}\sum_{i=1}^{6}\partial_{\nu}\phi^{i}
\partial^{\rho}\phi^{i}-{1\over 2}F_{\mu\nu}\sum_{i=1}^{6}
\partial_{\rho}\phi^{i}
\partial^{\rho}\phi^{i}\right)\;,
\ee
where $g_{\rm YM}$ is the SYM coupling, $F_{\mu\nu}$ is the 
$U(N)$ field strength and $\phi^i,\;i=1,\cdots, 6$ are the 
adjoint scalars. This deformed theory by
the operator ${\cal O}_{\mu \nu}$ can be extended to a complete 
theory with a simple description  which is noncommutative SYM theory.

On the other hand we could deform the theory 
by an other operator. In particular one can consider an operator 
with dimension 5 whose bosonic part is given
by \cite{BG}
\be
{\cal O}_{\mu}^{ij}=\frac{i}{g_{\rm YM}^2}{\rm Tr}\left(F_{\mu \nu}
\phi^{[i}D^{\nu}\phi^{j]}+\sum_{k=1}^6
D_{\mu}\phi^{[k}\phi^i\phi^{j]}\right)\; ,
\ee
where $D_{\mu}$ is the covariant derivative with respect to gauge 
field $A_{\mu}$ of field strength $F_{\mu\nu}$ and $[\cdots]$ is 
complete antisymmertization.

The deformed theory by this vector operator can be considered 
as low energy limit of {\it ``dipole theory''} which is a non-local 
field theory such that some fields of the theory have  dipoles with 
constant length. Such a theory was discussed in \cite{BG} in the 
context of T-duality in noncommutative geometry. In fact the 
T-duality of the gauge theory on a noncommutative torus can be 
extended to include fields with twist boundary condition. 

From string theory point of view this non-local theory can be 
obtained from worldvolume theory of a D-brane in the presence 
of non-zero B field with one leg along the brane worldvolume 
and the other along the transverse directions to the brane. 
This brane configuration 
was studied in \cite{{BG},{CDGR},{DGR},{BDGKR}} where the twisted
compactification was introduced. This twisted compactification 
leads us to introduce a new type of star product between
the fields at the level of effective field theory. This is the 
aim of this article to study the dipole field theory 
using gravity via AdS/CFT correspondence.

The plan of this paper is a following: In section 2 we will review 
the dipole field theory. In section 3 we shall construct the 
supergravity solutions corresponding to the D-branes in the 
presence of non-zero B field with
one leg along the brane worldvolume and the other transverse to it. 
We will show that for this D$_p$-brane system the worldvolume 
theory decouples from the gravity for $p\leq 5$. In section 4 we 
will use these supergravity  solutions to study dipole field 
theories via AdS/CFT correspondence.
In section 5 we shall consider the type II NS5-branes in the 
presence of RR field with 
one leg along the transverse directions of the branes and the 
others along the NS5-branes worldvolume. We then proceed to study 
M5-brane in the presence of C field with two legs along the
M5-brane worldvolume and the other transverse to the brane.
These solutions could provide new deformations 
of little string theory and (2,0) theory. In section 6 we 
shall compute the Wilson 
loops in the dipole gauge theory using the dual gravity description. 
The section 7 is devoted to conclusions and comments.

\section{Dipole field theory}

Dipole field theory can be thought as a generalization of ordinary 
field theory on a commutative space which is non-local theory 
and breaks Lorantz invariance. In order to turn the 
ordinary theory to non-local dipole theory one assigns, to every 
fields $\Phi_a$, a constant dipole vector with dipole length
$L_a$ and define the dipole star product as
\be
\Phi_a*\Phi_b(x)\equiv \Phi_a(x-{L_b\over 2})\Phi_b(x+{L_a\over 2})\;.
\ee
This product is associative if the vector assignment is
 additive, that is, $\Phi_a*\Phi_b$ has dipole vector $L_a+L_b$. 
Moreover, demanding $\Phi_a^{\dagger}*\Phi_a$ to be real fixes the 
dipole length of $\Phi_a^{\dagger}$ to be minus of that of $\Phi_a$. 
As a result, the gauge field has zero dipole length.  

In order to write an action for a dipole theory it is enough 
to replace the product of fields in the commutative action 
with the dipole star product. We note, however, that one 
should insert the proper dipole length for all fields. 
In particular, any term in the proposed action should have total 
zero dipole length \cite{JD}. 

Since the gauge field has zero dipole length we cannot have nontrivial
pure supersymmetric Yang-Mills theory. At least one needs to 
have a hyper-multiplet. In this case while
the vector-multiplet has zero dipole length, the hyper-multiplet 
could have a non-zero dipole. Therefore the maximal possible 
supersymmetric noncommutative
dipole theories, in dimensions less than seven, have 
8 supercharges. In 6-dimensions it is ${\cal N}=1$
SYM theory coupled to a hyper-multiplet and in 4-dimensions it is
 ${\cal N}=2$ coupled to an adjoint hyper-multiplet, and {\it etc.} 
In fact, these are the theories one can get from string theory. 

In the next section we are present the supergravity 
solution of D-brane in type II string theories in the presence of 
a non-zero B field with one leg along the worldvolume. 
These solutions would provide a string theory realization 
of dipole theories. Although in general such a configuration 
will break the supersymmetry completely, we shall consider
the cases which have maximal supersymmetry that are theories 
with 8 supercharges.

\section{The Supergravity solution}

In order to find supergravity solutions corresponding to the 
dipole field theories in various dimensions we start from type II 
string theories on a space that is $R^{9,1}$ modded out by the 
isometry
\be
{\cal U}: (x_0,\cdots, x_p,\{x_a\}_{a=p+1}^{9})\mapsto (
x_0,\cdots,x_p+2\pi \beta_p,
\{\sum_{b=p+1}^{9}O_{ab}x_{b}\}_{a=p+1}^{9})\; .
\label{ISO}
\ee
Here $O\in SO(9-p)$ is an orthogonal matrix. Actually this is 
the generalization of $p=3$ case considered in \cite{BDGKR}. 
The explicit from of $O$ is given by
$O=e^{\frac{2\pi i \beta_{p} M}{\alpha'}}$ where $M$ is a 
finite matrix of the Lie algebra $so(9-p)$ with dimension of length.

Now we want to probe these backgrounds with a system of $N$ 
D$_{(p-1)}$-branes with worldvolume directions $(x_0,\cdots,x_{p-1})$ 
and taking $\beta_p\rightarrow 0$ 
limit. When $M=0$ we can perform T-duality finding normal D$_p$-brane. 
On the other hand for the case of $M\neq 0$ performing T-duality  
will end up with a D$_p$-brane whose low energy effective theory 
is going to be a dipole theory \cite{BDGKR}.

To find the explicit supergravity solution of D$_p$-brane 
describing the dipole theory via the Maldacena conjecture \cite{MAL}, 
we start with the following type II supergravity solution describing
$N$ coincident extremal for $D_p$brane (in string frame)\cite{HS}
\bea
ds^2 &=& f ^{-\frac{1}{2}} (-dt^2 + dx^2_1 +\cdots+dx^2_p) + 
f ^{\frac{1}{2}}(
dx^2_{p+1} + \cdots + dx^2_9)\;, \cr &&\cr
e^{2 \Phi} &=& g^2_s f^{\frac{3-p}{2}}\; ,\;\;\;\;\;f=1+\frac{(2\pi)^{p-2}
c_pNg_sl_s^{7-p}}{r^{7-p}},
\;\;\;\;\; C_{0,\cdots, p}=-f^{-1},
\eea
where $c_p = 2^{7-2p} \pi^{\frac{9-3p}{2}} \Gamma (\frac{7-p}{2})$.

The supergravity solution of a D$_{(p-1)}$-brane smeared in $x_p$ 
direction can be obtained by wrapping the above supergravity 
solution on a circle along $x_p$ with radius 
$\beta_p$ and then T-dualizing in this direction. Using T-duality 
transformation \cite{GPR}, one finds
\bea
ds^2 &=& f ^{-\frac{1}{2}} (-dt^2 + dx^2_1 +\cdots+dx^2_{p-1}) + 
f ^{\frac{1}{2}}
(\frac{\alpha'^2}{\beta_p^2} d \hat{x}^2_p + dx^2_{p+1} +\cdots +
dx^2_9)\; , \cr &&\cr
e^{2 \phi} &=& \frac{\alpha'}{\beta_p^2} g^2_s f^{\frac{4-p}{2}}\; ,
\label{MET}
\eea
where $\hat{x}_p$ is now dimensionless angular coordinate with 
period $\hat{x}_p\sim \hat{x}_p+2\pi$.
 
By making use of the modded out isometry (\ref{ISO}) we can add 
a twist to the transverse directions $x_{p+1},\cdots, x_9$ as we 
go around the circle $x_p$. Under the action of $O$ in 
(\ref{ISO}) we have
\be 
\delta x_{a}=\sum_{b=p+1}^{9}\Omega_{ab}x_b d{\hat x}_p\;,\;\;\;\;\;
a=p+1,\cdots, 9\;,
\ee 
where $\Omega_{ab}$ is an element of the Lie algebra $so(9-p)$. 
As a result the metric (\ref{MET}) changes to 
\be 
ds^2 = f ^{-\frac{1}{2}} (-dt^2 + dx^2_1 +\cdots+dx^2_{p-1}) + f
^{\frac{1}{2}} \left( \frac{\alpha'^2}{\beta_p^2} d \hat{x}^2_p +
\sum_{a=p+1}^9 (dx_a - \sum_{b=p+1}^9 \Omega_{ab}x_b d \hat{x}_p 
\hspace{2mm} )^2
\right)\;.\nonumber 
\ee
Expanding this out, we get
\bea 
 ds^2 &=& f ^{-\frac{1}{2}}
(-dt^2 + dx^2_1 +\cdots+dx^2_{p-1}) \cr  && \cr &+& f ^{\frac{1}{2}}
\left( (\frac{\alpha'^2}{\beta_p^2} + X^T \Omega^T \Omega X) d
\hat{x}^2_p  + dX^T dX - 2 (dX^T \Omega X) d\hat{x}_p  \right)\; ,
\eea
where $X$ is a vector formed by the transverse directions, 
{\it i.e.} $X^T = (x_{p+1},\cdots, x_9)$.  

Finally, once again, we can apply T-duality on $ \hat{x}_p$ 
direction. Doing so, in the limit of $\beta_p\rightarrow \infty$ 
keeping $\beta_p \Omega=M$ fixed and setting 
$x_p=\beta_p {\hat x}_p$, one finds
\bea
ds^2 &=& f^{-\frac{1}{2}} \left( -dt^2 + dx^2_1 +\cdots+dx^2_{p-1} 
+ \frac{ \alpha'^2 d x^2_p}{ \alpha'^2 + r^2 n^T M^T M n} \right) 
\cr && \cr
&+& f^{\frac{1}{2}} \left( dr^2 + r^2 dn^T dn - 
\frac{r^4 (n^T M^T dn)^2}{\alpha'^2 + r^2 n^T M^T M n} 
 \right)\;, \cr &&\cr 
 e^{2\phi} &=&  \frac{\alpha'^2 g^2_s f^{\frac{3-p}{2}}}{{\a'}^2 +
 r^2 n^T M^T M n }\; ,\cr &&\cr 
\sum_{a=p+1}^9 B_{pa} dx_a &=& - \frac{r^2dn^TM n}{ \alpha'^2  + 
r^2 n^T M^T M n}\;,
\label{GenSol}
\eea
where $n$ is unit vector defined by $X=r n$ with $|n|^2=1$.

In general this background breaks the supersymmetry completely, 
though depending on the rank of matrix $M$ some supersymmetries
are left. We will return to this point in the next 
section. Another problem one has to be considered is the stability 
of the solution. It is not obvious if the solutions we found are 
stable. Nevertheless taking  matrix $M$ such that the solutions 
preserve some  amount of supersymmetries, as we will do in the next 
section, would hopefully lead to stable solutions.

On the other hand, given a general supergravity solution of a system 
of branes, it is not clear whether the solution would give a well-defined 
description of some field theory. In fact, we must check and see 
whether there is a well-defined field theory on the brane worldvolume 
which decouples from bulk gravity. To see this, one 
might calculate scattering amplitude for gravitons 
\cite{{GKT},{GK}}. To do this, we can compute the gravitons 
absorption cross section. If there is a limit (decoupling limit) 
where the gravitons absorption 
cross section vanishes, we have a field theory which decouples 
from the gravity. Alternatively, one can evaluate the 
potential that the gravitons feel because of the brane. Having a 
decoupled theory can be seen from the shape of the potential in 
the decoupling limit. Actually, for those branes which their 
worldvolume decouple from gravity, the potential develops an infinite 
barrier separating the space into two 
parts: bulk and brane. 
In this case the bulk's modes can not reach the brane because of this
infinite barrier, and the same for brane's modes. Therefore the theory 
decouples from the bulk. As an example, for D6-brane there is no such a 
barrier and therefore we expect not to have  a 7-dimensional gauge 
theory living on its worldvolume. This is also the case even with a 
non-zero B field \cite{AIO}. In following we are going to compute the 
potential that  the gravitons 
see because of the supergravity solution (\ref{GenSol}).

Let us perturb the metric of the background (\ref{GenSol}) by
\be 
g_{ij} = \bar{g}_{ij} + h_{ij} \hspace{1cm} i,j=0,\cdots,9, 
\ee
where by $\bar{g}_{ij}$ we denote the background metric (\ref{GenSol})
and $h_{ij}$ is the perturbation. We consider s-wave gravitons with 
momenta along the brane
\be
h_{ij} = \epsilon_{ij} h(r) e^{i k_{\mu} x^{\mu}} , \hspace{1cm} 
\mu = 0,\cdots, p \; .
\ee
We choose the gauge $h_{i\mu}k^{\mu}=0$ that keeps transversal 
gravitons. We will also choose the garvitons with polarization 
along the brane, {\it i.e.} $\epsilon_{ab}=0$ for $a,b=p+1,\cdots, 9$. 
Let $k_{\mu}=\omega \delta_{0,\mu}$. 
The other possibilities do not change the physical results as far as 
the decoupling is concerned. Using the linearized equations of motion 
of type II supergravity we find following equation for transverse 
gravitons
\be
\partial_{i} \left(\sqrt{-g} e^{- 2 \phi} g^{ij} \partial_{j}\Phi\right)=0
\ee
with $\Phi=h(r) e^{i k_{\mu} x^{\mu}}$. From this equation one can read 
the potential by writing it in the form of a Schr\"odinger-like equation. 
In general it might be difficult to do this for most general form of 
supergravity solution (\ref{GenSol}).
But we have done this for a particular form of matrix $M$ corresponding 
to the cases that preserve eight supercharges (see next section). 
In all cases the equation can be simplified and we get the 
following Schr\"odinger-like equation
\be
\partial_{\rho}^2 \psi(\rho) +V_p(\rho) \psi(\rho)=0\;,
\ee 
where 
\be V_p(\rho) = -\left(1+\frac{c_pNg_s(\omega l_s)^{7-p}}
{\rho^{7-p}}\right)+ \frac{(8-p)(6-p)}{4 \rho^2}
\ee 
with $\rho=\omega r$.

This potential is the same as once  found in \cite{AIO} for the 
ordinary D-branes as well as branes in the presence of B field. 
Therefore we conclude that we have a decoupled dipole theory 
living on the worldvolume of D$_p$-brane for $p\leq 5$.

In next section we shall study the dipole field theory living 
on the worldvolume of D-branes in the presence of B field with
one leg along the worldvolume and the other along the transverse 
directions to the brane  using corresponding supergravity solution
(\ref{GenSol}). We will only consider those
configurations that preserve 8 supercharges. 
Because of the supersymmetry we expect that the solution to be stable. 

\section{Supergravity description of dipole gauge theory}

In the previous section we have shown that the worldvolume theory 
on a system of D$_p$-brane in the presence of non-zero B field 
with one leg along the brane worldvolume decouples from 
the gravity for $p\leq 5$. Therefore this can 
provide a dual gravity description for 
dipole field theory \cite{BDGKR} via AdS/CFT correspondence. 

The decoupling limit is defined as a limit in which
$\alpha'\rightarrow 0$ and keeping the following quantities 
fixed
\be
u={r\over l^2_s}\;,\;\;\;\;\;\;{\bar g}_s=g_sl_s^{p-3}\; .
\ee
In this limit the supergravity solution (\ref{GenSol}) reads
\bea
l_s^{-2}ds^2 &=& (\frac{u}{R})^{\frac{(7-p)}{2}}
\left( -dt^2 + dx_1^2 +\cdots+ \frac{dx_p^2}{1+u^2 n^T M^T M n} \right)
 \cr &&\cr &+& (\frac{R}{u})^{\frac{7-p}{2}} \left(du^2 + u^2 dn^T dn 
- \frac{u^4(n^T M^T dn)^2}{1+u^2 n^T M^T M n} \right)\; , \cr &&\cr
e^{2 \phi} &=& {\bar g}_s^2 
\frac{(\frac{R}{u})^{(7-p)(3-p)/2}}{1+u^2 n^T M^T M n}
\; ,\cr &&\cr
\sum_{a=p+1}^9B_{pa}dn_a &=&-  
\frac{u^2 dn^TM n}{1 + u^2 n^T M^T M n}\; ,
\label{NEAR}
\eea
with
\be
(R)^{7-p} = 2^{7-2p} \pi^{(9-3p)/2} \Gamma (\frac{7-p}{2}) g^2_{YM} N\;,
 \;\;\;\;\;\;
g^2_{YM} = (2\pi)^{p-2} {\bar g}_s \; .
\ee
The effective dimensionless coupling constant in the corresponding
dipole field theory can be defined as following \cite{IMSY}
\be
g_{\rm eff}^2\sim g_{\rm YM}^2 N u^{p-3}\; .
\ee
For the cases we will be considering the scalar curvature of the metric
in eq. (\ref{NEAR}) has the behavior
\be
l_s^2{\cal R}\sim \frac{1}{g_{\rm eff}}\; .
\ee
Thus the perturbative calculation in 
dipole field theory can be trusted when $g_{\rm eff}\ll 1$, while when 
$g_{\rm eff}\gg 1$ the supergravity description is valid. 
We note also that the expression for dilaton in (\ref{NEAR}) 
can be recast to 
\be
e^{\phi}={1\over N}\;\frac{g_{\rm eff}^{(7-p)/4}}
{(1+u^2n^TM^TMn)^{1/2}}\;.
\label{DIL}
\ee
Keeping $g_{\rm eff}$ and $u^2n^TM^TMn$ fixed we see from 
(\ref{DIL}) that $e^{\phi}\sim 1/N$. Therefore the string
loop expansion corresponds to $1/N$ expansion of dipole gauge 
theory. Note also that in the supergravity description of 
dipole field theory $u^2n^TM^TMn$ plays the role of 
dipole deformation. At IR limit where $u^2n^TM^TMn\ll 1$
the dipole effects are small and the effective description of the 
worldvolume theory is in terms of a ordinary field theory.
In this regime the supergravity solutions (\ref{NEAR}) reduce
to the low energy background considered in \cite{IMSY}.

When the rank of $M$ is less than maximal, which of course
the case we are interested in, the quadratic form 
$n^TM^TMn$ has a locus of zeroes \cite{BDGKR}. Form the
supergravity solution (\ref{NEAR}) we see that locally on the
zero locus, solution is the same as ordinary D$_p$-brane
background \cite{IMSY}. 

The case of $p=3$  has been studied in \cite{{JD}, {BDGKR}}. 
Here, we will first review the D3-brane case and then we shall 
study the other branes.

\subsection{D3-brane} 

The theory on the worldvolume of the  $N$ coincident ordinary
D3-brane is ${\cal N}=4$ $SU(N)$ SYM theory. The ${\cal N}=4$ 
SYM theory in four dimensions has 6 real scalars in the 
representation {\bf 6} of $R$-symmetry group $SU(4)$ and 4 Weyl
fermions in the representation {\bf 4} of $SU(4)$. 
To construct the dipole theory we use the $R$-symmetry charges 
to determine the dipole vectors of the various fields.
For simplicity we assume that all dipoles are 
along $3^{rd}$ direction. The dipole moment can be given by 
vector $V^3\in su(4)$. Denoting the matrix representation
of $V^3$ by $U$ for representation {\bf 4} and by $M$ for 
representation {\bf 6}, the dipole vectors of fermions and 
scalars are given by eigenvalues of $U$ and $M$, respectively
\cite{BDGKR}.

In general the number of supersymmetries that are preserved by 
dipole theory is determined by the rank of $U$. It is completely 
broken for $U$ of rank 4. For $U$ with rank three, it has 
one zero eigenvalue and the theory has ${\cal N}=1$ supersymmetry. 
For $U$ of rank 2 there are two zero eigenvalues and the theory 
has ${\cal N}=2$ supersymmetry.    

Being a matrix in the representation {\bf 4} of $su(4)$ 
the eigenvalues of $U$ can be given by $\alpha_1, \alpha_2, 
\alpha_3$ and $-\alpha_1-\alpha_2-\alpha_3$. 
Therefore the most general form of matrix $M$ can be 
cast to the following form
\be
M=\pmatrix{0&\alpha_1+\alpha_2&0&0&0&0 \cr -\alpha_1-\alpha_2 
&0&0&0&0&0\cr 0&0&0&\alpha_1+\alpha_3&0&0 \cr 0&0& -\alpha_1
-\alpha_3 &0&0&0\cr 0&0&0&0&0&\alpha_2+\alpha_3 \cr 0&0&0&0& 
-\alpha_2-\alpha_3 &0}\;.
\ee
This form of matrix $M$ breaks all supersymmetries. 
On the other hand for $\alpha_1=0$ we left with 4 
supercharges. For $\alpha_1=\alpha_2=0$ we find a configuration 
with 8 supercharges. Setting $\alpha_3=L$, the matrix $M$ reads
\be
M=\pmatrix{0&0&0&0&0&0 \cr 0 &0&0&0&0&0\cr
0&0&0&L&0&0 \cr 0&0& -L &0&0&0\cr
0&0&0&0&0&L \cr 0&0&0&0& -L &0}\; .
\ee
According to the definition of $M$ \cite{BDGKR}, we have 
four real scalars with dipole lengths $\pm 2\pi L $ and two 
Weyl fermions with dipole lengths $\pm \pi L$. These are 
content of an ${\cal N}=2$ hyper-multiplet. The fields in the 
${\cal N}=2$ vector-multiplet have no dipole and this is the 
reason why we have 8 supercharges.

Plugging the above form of $M$ into (\ref{NEAR}), one finds
\bea
l_s^{-2}ds^2&=& (\frac{u}{R})^2 \left( -dt^2 + dx_1^2 + dx_2^2+ 
\frac{dx_3^2}{1+u^2L^2\sin^2\theta_1 \sin^2 \theta_2} \right)
 \hspace{4cm}\cr &&\cr &+& (\frac{R}{u})^2 \left( du^2 + u^2 d 
 \Omega_5^2 - u^4 L^2\sin^4 \theta_1 \sin^4\theta_2
 \frac{( a_3d \theta_3 + a_4 d\theta_4+a_5 d\theta_5)^2 } 
 {1+u^2L^2\sin^2\theta_1 \sin^2 \theta_2} \right),\cr &&\cr
e^{2 \phi} &=&  \frac{{\bar g}_s^2}{1+u^2L^2\sin^2\theta_1 \sin^2 
\theta_2}\;,
\cr &&\cr
\sum_{a=3}^5B_{3\theta_a}d\theta_a&=&-
\frac{ u^2L\sin^2\theta_1\sin^2\theta_2 } 
{1+u^2L^2\sin^2\theta_1 \sin^2 \theta_2}(a_3d \theta_3 +
 a_4 d\theta_4+a_5 d\theta_5)\; ,
\eea
where $\theta_{1}\cdots\theta_{5}$ are the angular coordinates
parameterizing sphere $S^5$ transverse to  the D3-brane, and 
\be
a_3= \cos\theta_4 , \hspace{12mm} a_4=-\sin{\theta_3}\cos{\theta_3}
\sin{\theta_4},\hspace{12mm} a_5=\sin^2 \theta_3 \sin^2 \theta_4\;.
\ee
The locus of zeros is given by equation $\sin\theta_1\sin\theta_2=0$ 
where the gravity solution is the same as the case without B field.
The phase structure of this theory was studied in \cite{{JD}}.

\subsection{D5-brane }

As we showed in the previous section, since D6-brane does not 
decouple from the gravity, the next example we are going to 
consider is D5-brane. The low energy effective theory on the 
worldvolume 
of the  ordinary D5-brane is 6-dimensional gauge theory with 
16 supercharges. It has $SO(4)$ $R$-symmetry.
One can use $R$-symmetry charges to make a dipole theory out of the 
ordinary one. In order to have a 6-dimensional supersymmetric 
dipole theory we need to separate the ${\cal N}=2$ vector-multiplet 
to an ${\cal N}=1$ vector-multiplet plus a hyper-multiplet. 
Now we can assign to the fields in the hyper-multiplet a dipole 
vector and therefore we get ${\cal N}=1$
6-dimensional supersymmetric dipole gauge theory. 
The dipole vectors of the scalars in the hyper-multiplet can be 
obtained from the eigenvalues
of matrix $M$ which can be cast to the following form 
\be
M=\pmatrix{0&L&0&0\cr -L&0&0&0\cr 0&0&0&L\cr 0&0&-L&0}\;.
\label{SM5}
\ee
Plugging the above matrix in (\ref{NEAR}) we get
\bea
l_s^{-2}ds^2 &=& \frac{u}{R}\left( -dt^2 + dx_1^2 
+\cdots+ \frac{dx_5^2}{1+u^2L^2} \right) \cr &&\cr  
 &+& \frac{R}{u} \left( du^2 + u^2 d \Omega_3^2 - \frac{u^4L^2}
{1+u^2L^2}(a_1d\theta_1 
 +a_2 d\theta_2 + a_3d \theta_3)^2 \right)\; ,\cr &&\cr
e^{2\phi}&=&{\bar g}_s^2\frac{\left({u\over R}\right)^2}{1+u^2 L^2}\;,
\cr &&\cr 
\sum_{a=1}^3B_{5\theta_a}d\theta_a&=&-
\frac{u^2L  } {1+u^2L^2}(a_1d \theta_1 +
 a_2 d\theta_2+a_3 d\theta_3)\;,
\label{DD5} 
\eea
where $\theta_i$'s are angular coordinates parameterizing sphere 
$S^{3}$ transverse to the brane, and
\be
a_1= \cos\theta_2,\;\;\;\;\; a_2=-\sin \theta_1 \cos 
\theta_1 \sin \theta_2,
\;\;\;\;\;a_3=\sin^2 \theta_1 \sin^2 \theta_2\;.
\label{ASM5}
\ee

To study the phase structure of the D5-brane theory, we realize
that there are three energy scales which involved. The first one
is the energy scale in which the dipole effects become important, 
that is $u\sim 1/L$. The second one is the energy in which the 
curvature is of order one $u\sim 1/\sqrt{{\bar g}_s N}$ and finally the
dilaton is of order one while the dipole effects are
negligible at $u\sim \sqrt{N/{\bar g}_s}$. 
Using these information we recognize three
different phase structures for the theory as following:

\begin{enumerate}
	\item Suppose $\sqrt{N/{\bar g}_s}\ll 1/ L$. 
At extreme IR regime the theory can be described by 
perturbative SYM theory. As we increase the energy
at the scale $u\sim 1/\sqrt{{\bar g}_s N}$ the curvature is of 
order one and then we have to change our description. For 
$1/\sqrt{{\bar g}_s N}\ll u\ll\sqrt{N/{\bar g}_s}$ the good 
description is given by gravity solution of D5-brane. For 
the energy of order $\sqrt{N/{\bar g}_s}$ the dilation is of order 
one, thus by passing this energy
we need to change our gravity description using S-duality. 
Therefore for $u\gg \sqrt{N/{\bar g}_s}$ NS5-brane description 
is valid. As we keep increasing the energy the effects of B field 
become important, and eventually at UV we will have
a theory which could be considered as a new deformation of little 
string theory. We will return to this solution in the next section.	
	
	\item When $1/\sqrt{{\bar g}_s N}\ll 1/L \ll\sqrt{N/{\bar g}_s}$. 
The IR regime is again described by perturbative SYM theory. 
For $ 1/\sqrt{{\bar g}_s N}\ll u\ll 1/L$ the curvature and dilation 
are small. Moreover the effects of B field are still negligible. 
So, the theory can be described by gravity solution of ordinary 
D5-brane. As we pass the scale $1/L$ the effects of B field become 
important. From the gravity solution (\ref{DD5}) we see that the 
dilation remains small all the time and therefore the good 
description is given by gravity solution
of D5-brane in the presence of B field (\ref{DD5}).   
	
	\item For the case of $1/L\ll  1/\sqrt{{\bar g}_s N}$, the extreme 
IR regime is described by perturbative SYM theory. At the scale of 
$1/L$ the curvature is still large but the dipole effects become 
important and thus the good description in
region $1/L\ll u \ll 1/\sqrt{{\bar g}_s N}$
is given by perturbative dipole gauge theory. By passing the 
energy scale $1/\sqrt{{\bar g}_s N}$ the curvatures becomes small 
while the effects of B field are important. Therefore for 
energy $u\gg 1/\sqrt{{\bar g}_s N}$
the good description is given by gravity solution
of D5-brane in the presence of B field (\ref{DD5}).
	
\end{enumerate}
 
A new feature of 6-dimensional dipole field theory is that there 
is a regime in which we have perturbative dipole gauge theory 
description. We note that this is not the
case when we have noncommutative brane \cite{AOJ}. The reason 
is that in the noncommutative case the scale in which the effects 
of noncommutativity are important is written in terms of 
$g_{\rm eff}$ and it is not possible to have $g_{\rm eff}\ll 1$
while the noncommutative effects remain important. In other words, 
in the phase diagram we do not have a range of energy where the 
perturbative noncommutative gauge theory is valid \cite{AOJ}.

\subsection{D4-brane}
Having the supergravity solution for D5-brane with 
8 supercharges, we can easily find a solution for D4-brane 
which preserves 8 supercharges as well. Here the matrix 
$M$ representing the dipoles can be written as 
\be
M=\pmatrix{0&0&0&0&0\cr 0&0&L&0&0\cr 0&-L&0&0&0\cr 0&0&0&0&L
\cr 0&0&0&-L&0}\;.
\label{MD4}
\ee
Plugging this matrix into the supergravity solution (\ref{NEAR}), 
one finds
\bea
l_s^2ds^2&=& (\frac{u}{R})^{3/2}\left( -dt^2 + dx_1^2 + \cdots+dx_3^2+ 
\frac{dx_4^2}{1+u^2L_{\rm eff}^2} \right) \cr &&\cr 
 &+& (\frac{R}{u})^{3/2} \left( du^2 + u^2 d \Omega_3^2 - 
 \frac{u^4 L_{\rm eff}^2} {1+u^2L_{\rm eff}^2} 
(a_2d\theta_2 +a_3 d\theta_3 + a_4d \theta_4)^2 \right)\;,\cr &&\cr
e^{2 \phi} &=&{\bar g}_s^2  \frac{\left({u\over R}\right)^{3/2}}
{1+u^2L_{\rm eff}^2 }\;,
\cr &&\cr
\sum_{a=2}^4B_{4\theta_a}d\theta_a&=&-
\frac{ u^2L_{\rm eff}} 
{1+u^2L_{\rm eff}^2}(a_2d \theta_2 +
 a_3 d\theta_3+a_4 d\theta_4)\; ,
\eea 
where $\theta_1,\cdots,\theta_4$ are angular coordinates 
parameterizing the
sphere $S^4$ transverse to the brane, and
\be
a_2= \cos\theta_3, \hspace{1cm} a_3=-\sin \theta_2 \cos \theta_2 
\sin \theta_3,
\hspace{1cm} a_4=\sin^2 \theta_2 \sin^2 \theta_3\; .
\label{AD4}
\ee
Moreover the effective dipole vector is defined by 
$L_{\rm eff}=L\sin\theta_1$. As we see
there are a locus of zeros given by $\sin\theta_1$ in which the 
effective dipole is zero. We note that the effective dipole
vector seen from supergravity is smaller than the naively
expected dipole vector $L$. This can be interpreted as
due to strong interactions. This is very similar to what
we have in the noncommutative gauge theory case \cite{MR}.

The three energy scales that involved in parameterizing the phase
diagram of the theory are as following. At $u\sim 1/{\bar g}_sN$ 
the curvature is of order one. The dipole effects become important
at scale $u\sim 1/L_{\rm eff}$, while the dipole 
effects are negligible the dilaton is of order 
one at $u\sim N^{1/3}/{\bar g}_s$. 
The same as D5-brane case we 
distinguish three different phase structures of the 
theory as following:
\begin{enumerate}
\item When $1/L_{\rm eff}\gg N^{1/3}/{\bar g}_s$ the theory at IR
limit is described by perturbative 5-dimensional SYM theory.
In flow from low energy to high energy we need to change our description
from perturbative SYM theory to gravity description at energy 
$u\sim 1/{\bar g}_sN$ where the curvature becomes small. In fact the
good description of the theory in energy range $1/{\bar g}_sN\ll u
\ll N^{1/3}/{\bar g}_s$ is given by gravity solution of D4-brane. 
As we pass energy $u\sim N^{1/3}/{\bar g}_s$ 
the dilaton gets large and we have to lift the theory to eleven 
dimensional SUGRA and the uplifted D4-brane solution is a good 
description. Eventually at the scale $u\sim 1/L_{\rm eff}$ the 
B field effects become important.
Because of these effects the dilaton becomes small at $u\sim
{\bar g}_s^3/NL_{\rm eff}^4$ and therefore the good
description at UV will be given by the supergravity solution 
of D4-brane in the presence of B field with one leg along the 
worldvolume and the other transverse to it. 

\item For the case of $1/{\bar g}_s N\ll 1/L_{\rm eff}\ll N^{1/3}/{\bar g}_s$
the IR limit is described by perturbative 5-dimensional SYM theory. As 
we increase the energy we have to change our description from perturbative
SYM theory to the gravity description at scale $u\sim 1N^{1/3}/{\bar g}_s$.
At scale $1/L_{\rm eff}$ the B field effects become important while
the dilaton remains small all the time. As a result the good 
description at UV is given by type IIA supergravity solution
of D4-brane in the presence of B field with one leg along the brane and
the other transverse to it. 

\item For $1/L_{\rm eff}\ll 1/{\bar g}_s N$ at extreme IR limit the 
perturbative
SYM theory is valid. As we increase the energy at scale 
$u\sim 1/L_{\rm eff}$ the
good description is given by perturbaive dipole gauge theory. 
This description is valid till we reach the energy 
$u\sim 1/{\bar g}_s N$ where the curvature is
of order one and we have to change our description from 
perturbative dipole gauge
theory to gravity solution of D4-brane in the presence of B field. Since the 
dilation remains small all the time this is also the valid description at
UV limit.
\end{enumerate}

Again, in comparison with noncommutative field theory we see that
for dipole field theory there is a range of energy in which the
good description is given by perturbative dipole gauge theory.

\subsection{D2-brane}
One can generalize the previous solutions to the supergravity 
solution of D2-brane
in the presence of B field with one leg along the worldvolume 
that preserves 8 supercharges. The solution is
\bea
l_s^2ds^2 &=& (\frac{u}{R})^{5/2} \left( -dt^2 + dx_1^2 + \frac{dx_2^2}
{1+u^2L_{\rm eff}^2} \right)\cr &&\cr 
 &+& (\frac{R}{u})^{5/2} \left( du^2 + u^2 d \Omega_6^2 - 
 \frac{u^4 L_{\rm eff}^2 } 
 {1+u^2L_{\rm eff}^2}
( a_{\theta_4}d \theta_4 +a_{\theta_5} d\theta_5+a_{\theta_6} 
d\theta_6)^2 \right),
\cr &&\cr
e^{2 \phi} &=& {\bar g}_s^2 
\frac{\left({R\over u}\right)^{5/2}}
{1+u^2L_{\rm eff}^2}\;,
\cr &&\cr
\sum_{a=4}^6B_{2\theta_a}d\theta_a&=&-
\frac{ u^2L_{\rm eff} } 
{1+u^2L_{\rm eff}^2}(a_4d \theta_4 +
 a_5 d\theta_5+a_6 d\theta_6)\; ,
\eea
where $\theta_1,\cdots,\theta_6$ are angular coordinates parameterizing the
sphere $S^6$ transverse to the brane, and
\be
a_4= \cos\theta_5 , 
\hspace{12mm} a_5=-\sin{\theta_4}\cos{\theta_4}\sin{\theta_5},
\hspace{12mm} a_6=\sin^2 \theta_4 \sin^2 \theta_5\; .
\ee
The effective dipole vector is defined by $L_{\rm eff}=L\sin\theta_1
\sin \theta_2\sin \theta_3$. The dipole vector could be zero as locus where
$\sin\theta_1\sin \theta_2\sin \theta_3$ vanishes. In this locus 
the gravity solution
is locally the same as ordinary D-brane solution \cite{IMSY}.

The three scales which play role in studying of the phase structure 
of D2-brane are as following. The curvature is of order one at 
$u\sim {\bar g}_s N$. The dipole effects are negligible below the
energy of order $u\sim 1/L_{\rm eff}$ and for the case where
the dipole effects are negligible the dilaton is of order one
at $u\sim {\bar g}_s/ N^{1/5}$. Using these information, one
can proceed to analyze the phase structure of the theory.
\begin{enumerate}
\item When $1/L_{\rm eff}\gg {\bar g}_sN$ the good description at
UV is given by perturbative 3 dimensional dipole gauge theory 
with 8 supercharges. In flow from high energy to low energy the
dipole effects become negligible at $u\sim 1/L_{\rm eff}$ and 
therefore at energy range ${\bar g}_sN\ll u\ll 1/L_{\rm eff}$
the good description is given by perturbative SYM theory where
we have supersymmetry enhancement. At energy scale
$u\sim {\bar g}_sN$ the curvature is of order one and we need
to change our description from perturbative SYM theory to
gravity solution of D2-brane. The gravity solution of D2-brane
is a good description till we reach scale $u\sim {\bar g}_s/ N^{1/5}$
where the dilaton grows and we need to lift the theory to eleven
dimensional SUGRA and thus the good description is given
by uplifted D2-brane. Finally at IR regime the theory flows to
a fixed point which is given by ${\cal N}=2$ SCFT in three 
dimensions.
\item When ${\bar g}_s/ N^{1/5}\ll 1/L_{\rm eff}\ll {\bar g}_sN$
the theory is described by perturbative 3-dimensional supersymmetric
dipole gauge theory for
energy above ${\bar g}_sN$. At this energy the curvature is of 
order one and therefore we have to change our description at 
this scale. In fact the good description of theory is given in terms
of supergravity solution of D2-brane in the presence of B field
in energy range $1/L_{\rm eff}\ll u \ll{\bar g}_sN$ while 
below the scale $1/L_{\rm eff}$ the B field effects are negligible
and the theory is described by D2-branes solution. At energy
$u\sim {\bar g}_s/ N^{1/5}$ the dilaton is of order one and 
thus below this energy the uplifted D2-brane solution is a 
good description. Eventually the theory
reach its fixed point at IR where the good description is
given by ${\cal N}=2$ SCFT in three dimensions.
\item In the case of $1/L_{\rm eff}\ll {\bar g}_s/ N^{1/5}$ the
UV regime is described by perturbative supersymmetric dipole  
gauge theory. As we decrees the energy the effective 
dimensionless gauge coupling
grows at $\sim {\bar g}_sN$ where we have to change our description 
to the gravity description. The theory is described by supergravity 
solution of D2-brane in the presence of B field until we reach
the energy scale $u\sim {\bar g}_s^{5/9}N^{1/9}/L_{\rm eff}^{4/9}$
where the dilaton start growing and the good description will
be given by uplifted D2-brane in the presence of C field. The dipole 
effects become negligible at $1/L_{\rm eff}$ and finally the IR limit 
is described by the fixed point of the theory which is 
${\cal N}=2$ SCFT in three dimensions.

\end{enumerate}

\subsection{D1-brane}

Using the same method as before the supergravity solution of 
D1-brane in the presence of a B field with one leg along brane and 
the other transverse to it which preserves
8 supercharges can be read from (\ref{NEAR}), that is
\bea
l_s^2ds^2&=& (\frac{u}{R})^{3} \left( -dt^2 + \frac{dx_1^2}
{1+u^2L_{\rm eff}^2} \right) 
\cr &&\cr 
&+& (\frac{R}{u})^{3} \bigg{(} du^2 + u^2 d \Omega_7^2-
\frac{ u^4 L_{\rm eff}^2 } 
{1+u^2L_{\rm eff}^2}( a_5d \theta_5 +a_6 d\theta_6+
a_{7} d\theta_7)^2\bigg{)},
\cr &&\cr
e^{2 \phi} &=& {\bar g}_s^2 
\frac{\left({R\over u}\right)^{6}}
{1+u^2L_{\rm eff}^2}\;,
\cr &&\cr
\sum_{a=5}^7B_{1\theta_a}d\theta_a&=&-
\frac{ u^2L_{\rm eff}} 
{1+u^2L_{\rm eff}^2}
(a_5d \theta_5 + a_6 d\theta_6+a_7 d\theta_7)\; ,
\label{DD1}
\eea 
where $\theta_1,\cdots,\theta_7$ are angular coordinates parameterizing the
sphere $S^7$ transverse to the brane, and 
\be
a_{5}= \cos\theta_6 , \hspace{12mm} a_{6}=-\sin{\theta_5}
\cos{\theta_5}\sin{\theta_6},
\hspace{12mm} a_{7}=
\sin^2 \theta_5 \sin^2 \theta_6\;.
\ee
The effective dipole vector is defined by $L_{\rm eff}=L\sin\theta_1
\sin \theta_2\sin \theta_3\sin \theta_4$. 
The dipole vector could be zero in locus where
$\sin\theta_1\sin \theta_2\sin \theta_3\sin \theta_4$ vanishes. 
In this locus the gravity solution
is locally the same as ordinary D-brane solution \cite{IMSY}.

As it can be seen from the supergravity solution (\ref{DD1}) the curvature is
of order one at $u\sim g_{\rm YM}\sqrt{N}$. The dipole effects are important
at the scale $u\sim 1/L_{\rm eff}$. While the dipole effects are negligible
the dilaton is of one at $u\sim g_{\rm YM}N^{1/6}$. These are the scales
that involved in phase structure of D1-brane. By making use of these 
information
we can work out the phase structure of the theory as following.
\begin{enumerate}
\item For the case of $1/L_{\rm eff}\gg g_{\rm YM}\sqrt{N}$ the good 
description at UV is given by perturbative (1+1)-dimensional supersymmetric
dipole gauge theory. As we flow from high energy to low energy the dipole
effects become negligible at $u\sim  1/L_{\rm eff}$. In fact for
$g_{\rm YM}\sqrt{N}\ll u\ll 1/L_{\rm eff}$ the SYM theory
in (1+1)-dimensional can be trusted. As we approach the low
energy the type IIB supergravity solution (\ref{DD1}) can be
trusted in the region $g_{\rm YM}N^{1/6}\ll u\ll
g_{\rm YM}\sqrt{N}$. In the region $u\ll g_{\rm YM}N^{1/6}$
the dilaton is large and we need the S-dual picture. Using
S-duality the solution (\ref{DD1}) maps to the fundamental
string solution. In IR limit
the string coupling vanishes while the curvature in the new
string unit behaves as $l_{s}^2{\cal R}\sim g_{\rm YM}^2/u^2$.
This means that the supergravity description of 
fundamental string breaks down for small $u$ and in fact the
IR limit is a trivial orbifold $(R^{8N}/S_N)$ conformal
field theory \cite{IMSY}.
\item When $g_{\rm YM}N^{1/6}\ll 1/L_{\rm eff}\ll 
g_{\rm YM}\sqrt{N}$ the perturbative (1+1)-dimensional dipole
gauge theory is valid for $u\gg g_{\rm YM}\sqrt{N}$. While for
for $1/L_{\rm eff}\ll u\ll g_{\rm YM}\sqrt{N}$ the curvature 
and dilaton are small and the supergarvity solution of D1-brane
with B field is a good description. For $g_{\rm YM}N^{1/6}\ll u
\ll 1/L_{\rm eff}$ the gravity is still a good description but
the B field effects are negligible. For the energy $u\ll 
g_{\rm YM}N^{1/6}$ the situation is the same as previous case.
	
\item When $1/L_{\rm eff}\ll g_{\rm YM}N^{1/6}$ we can trust
perturbative (1+1)-dimensional dipole
gauge theory for $u\gg g_{\rm YM}\sqrt{N}$. In the region of
$(g_{\rm YM}^6N)^{1/8}/L_{\rm eff}^{1/4}\ll u\ll _{\rm YM}\sqrt{N}$
the curvature is small and dilaton is small too and the
good description is given by supergravity solution (\ref{DD1}). 
For the energy of $u\ll g_{\rm YM}^6N)^{1/8}/L_{\rm eff}^{1/4}$ the
dilaton is large while the B field effects are important and therefore
we need to use the S-dual picture which maps the solution (\ref{DD1}) to the
supergravity solution of fundamental string in the presence of RR 2-form
with one leg along the string and the other transverse to it. The B field 
effects becomes negligible for $u\ll 1/L_{\rm eff}$ where the gravity 
solution of fundamental string is valid. Eventually for $u\ll g_{\rm YM}$
the supergravity description of fundamental string breaks down and in 
fact the IR limit is a trivial orbifold $(R^{8N}/S_N)$ conformal
field theory.
	 
\end{enumerate}

\section{Fivebranes}
In this section we shall discuss possible dipole deformation of 
NS5-branes and M5-branes worldvolume theories.

\subsection{NS5-brane}

Now we want to  study the theory on the type II NS5-branes in the presence
of different RR fields with one leg along the transverse direction to the
NS5-branes and the others along the worldvolume. The supergravity
solution of NS5-branes in the presence of RR $(6-p)$-form, for
$p=0,\cdots, 4$, with one leg along the
transverse directions and $(5-p)$ legs along the NS5-branes worldvolume 
is given by
\bea
ds^2&=&(1+u^2n^TM^TMn)^{1/2}\left[\sum_{i=0}^{p}dx_i^2+
\frac{\sum_{j=p+1}^5dx_j^2}
{1+u^2n^TM^TMn}\right. \cr &&\cr 
&+&\left. g_s^2l_s^4f\left(du^2+u^2dn^Tdn-\frac{u^4(n^TM^Tdn)^2}
{1+u^2n^TM^TMn}\right)\right]\;,\cr &&\cr
e^{2\phi}&=&g_s^2(1+u^2n^TM^TMn)^{(p-2)/2}f\;,\cr &&\cr
\sum_{a=6}^{9}C_{(p+1)\cdots 5a}dx_a&=&-\frac{u^2 dn^TMn}
{1+u^2n^TM^TMn}\;,
\eea
where $dx^2_0=-dt^2$, $C_{(p+1)\cdots 5a}$ is $(6-p)$-form and 
$f=1+c_5N/g_s^2l_s^2u^2$. The energy coordinat $u$ is
related to the radial coordinat $r$ by $u=r/g_sl_s^2$.
A way to find these solutions 
is to start with 
type IIB D5-branes in the presence of B field with one leg along
the worldvolume. Then using S-duality we can find the supergravity solution
of type IIB NS5-branes in the presence of RR 2-form with one leg along 
the worldvolume. Other solutions can be obtained by T-duality. Under
S-duality we have
\be
l^2_s\rightarrow g_s l_s^2\;,\;\;\;\;\; g_s\rightarrow {1\over g_s}\;.
\ee
Therefore the decoupling limit of above supergravity solutions can be 
defined as the limit $g_s\rightarrow 0$, keeping the following quantities 
fixed
\be
u={r\over g_sl_s^2}\;,\;\;\;\;\;\; l_s={\rm fixed}\;.
\ee
In this limit, setting $M$ as (\ref{SM5}), the above supergravity solution
becomes
\bea
ds^2&=&(1+u^2L^2)^{1/2}\left[\sum_{i=0}^{p}dx_i^2+
\frac{\sum_{j=p+1}^5dx_j^2}
{1+u^2L^2}\right. \cr &&\cr 
&+&\left. {Nl_s^2\over u^2}\left(du^2+u^2d\Omega_3-\frac{u^4L^2}
{1+u^2L^2}(a_1d\theta_1+a_3d\theta_1+a_3d\theta_3)^2\right)\right]\;,\cr &&\cr
e^{2\phi}&=&\frac{N}{l_s^2 u^2}(1+u^2L^2)^{(p-2)/2}\;,\cr &&\cr
\sum_{a=6}^{9}C_{(p+1)\cdots 5\theta_a}d\theta_a&=&-\frac{u^2L}
{1+u^2L^2}(a_1d\theta_1+a_3d\theta_1+a_3d\theta_3)\;,
\eea
where $a_i$'s are defined the same as (\ref{ASM5}). This solution can be 
thought as a new deformation of little string theory.
The curvature of the metric is given by
\be
l_s^2{\cal R}\sim \frac{1}{N}\;\frac{1}{(1+u^2L^2)^{1/2}}\;.
\ee
Therefore for large $N$ the curvature is small and one can trust the 
supergravity description. For $uL\gg 1$ the gravity approximation
can be trusted for finite $N$. 

As an application let us to
calculate the absorption cross section of polarized
gravitons. In general one can show that in these 
backgrounds the scattering potential for a graviton polarized along
the brane directions is
\be
V(\rho)=-1+({3\over 4}-N\omega^2 l_s^2){1\over \rho^2}\; ,\;\;\;\;
\ee
where $\rho=r\omega$ and $\omega$ is the energy of incoming waves.
Therefore we see that after the decoupling limit the absorption cross section
can be nonzero only for waves with energy $\omega^2$ larger than 
$\sim {1\over
Nl_s^2}$.  Essentially the same effect appears in the little string 
theory and
one can see that the theory has a mass gap of order $M_{\rm gap}^2
\sim {1\over
Nl_s^2}$ \cite{MS} which is exactly the same as little string theory.   
In other words, the mass gap of the theory is independent of the
dipole vector. The mass gap in deformed little string theory has
been also studied in \cite{{ALI},{BR}}.

\subsection{M5-branes}

To find an eleven dimensional supergravity solution corresponding
to the M5-branes in the presence of a C field with two legs along
the worldvolume and one leg transverse to it, we can start from
$D_4$ brane solution and then lifting it to 11-dimensional 
supergravity and sending the radius of $11th$ direction to infinity,
$R_{11}\rightarrow \infty$. In this limit setting $R_{11}M={\bar M}$,
one finds  :
\bea
ds_{11}^2&=& (1+{r^2\over l_p^6}n^T{\bar M}^T{\bar M}n)^{1/3}\bigg{[} 
f^{-1/3}\left( -dt^2+\cdots+dx_3^2+ \frac{dx_4^2+dx_5^2}{1+
{r^2\over l_p^6}n^T{\bar M}^T{\bar M}n}\right)\cr&&\cr
&+& f^{2/3}
\bigg{(}dr^2+r^2dn^Tdn -\frac{{r^4\over l_p^6}(n^T{\bar M} dn)^2} 
{1+{r^2\over l_p^6}n^T{\bar M}^T{\bar M}n} \bigg{)}\bigg{]}\;,\cr &&\cr
\sum_{a=2}^4C_{45a}dx^a&\sim& -\frac{{r^2\over l_p^6}dn^T{\bar M} n} 
{1+{r^2\over l_p^6}n^T{\bar M}^T{\bar M}n}\; ,
\eea
where
\be
f=1+\frac{\pi N l_p^3}{r^3}\; .
\ee
The decoupling limit of the theory is defined as a limit where 
$l_p\rightarrow 0$ keeping $u={r\over l_p^3}$ fixed. In this limit,
setting ${\bar M}$ as
\be
{\bar M}=\pmatrix{0&0&0&0&0\cr 0&0&{\bar L}&0&0\cr 0&-{\bar L}&0&0&0\cr
0&0&0&0&{\bar L}\cr 0&0&0&-{\bar L}&0}\;,
\ee 
the above supergravity solution reads
\bea
l_s^2ds^2&=& (1+u^2L_{\rm eff}^2)^{1/2}\bigg{[}\frac{u}{(\pi N)^{1/3}}
\left( -dt^2 + dx_1^2 + \cdots+dx_3^2+ 
\frac{dx_4^2+dx_5^2}{1+u^2L_{\rm eff}^2} \right) \cr &&\cr 
 &+& \frac{(\pi N)^{2/3}}{u^2} \left( du^2 + u^2 d \Omega_3^2 - 
 \frac{u^4 L_{\rm eff}^2} {1+u^2L_{\rm eff}^2} 
(a_2d\theta_2 +a_3 d\theta_3 + a_4d \theta_4)^2 \right)\bigg{]},\cr &&\cr
\sum_{a=2}^4C_{45\theta_a}d\theta_a&=&-
\frac{ u^2L_{\rm eff}} 
{1+u^2L_{\rm eff}^2}(a_2d \theta_2 +
 a_3 d\theta_3+a_4 d\theta_4)\; ,
 \label{M5D}
\eea 
where $\theta_1,\cdots,\theta_4$ are angular coordinates 
parameterizing the
sphere $S^4$ transverse to the brane and $a_i$'s are given
by (\ref{AD4}). The effective ``{\it discpole}'' is also defined by 
$L_{\rm eff}={\bar L} \sin\theta_1$ where $L$ has dimension of 
$({\rm length})^2$.

The curvature of the metric in (\ref{M5D}) is
\be
l_p^2{\cal R}\sim {1\over N^{2/3}}\; \frac{1}{(1+u^2L_{\rm eff}^2)^{1/3}}\;.
\ee
Thus we can trust the supergravity solution for large $N$. This solution
can provide a dual description of a deformation of (2,0) theory. The dipole
deformation of (2,0) has also considered in \cite{DGR} as 
``{\it discpole theory}''.

\section{Wilson loop}

In this section we use  dual gravity description of dipole gauge 
theory to compute 
Wilson loop for different brane theories. According to the AdS/CFT 
correspondence the Wilson loop of the gauge theory can be computed in 
dual string theory description by evaluating the partition function
of string whose worldsheet is bounded by the loop \cite{{MAL2},{REY}}. 
In the supergravity approximation the dominant contribution comes from the
minimal two dimensional surface bounded by the loop. 
The expectation value of Wilson loop is
 \be
 \left\langle W(C) \right\rangle \sim e^{-S}\;,
\ee
where S is string action evaluated on the minimal surface bounded by loop C.

Now we would like to compute the Wilson
loop in the dipole gauge theories living on the worldvolume 
of D$_p$-brane using supergravity solution (\ref{NEAR}). The string action
is given by
\be
S= \frac{1}{2 \pi l_s^2} 
\int d \tau  d \sigma \sqrt{-\det \left( G_{\mu \nu} 
\partial_i X^{\mu} \partial_j X^{\nu} \right)}\;. 
\label{Wilson}
\ee
We parametrized the string configuration by $\tau = t, \sigma = u$ and 
$x_p = x(u)$. In this parameterization, using the supergravity
solution (\ref{NEAR}), the string action (\ref{Wilson}) reads 
\be
S= \frac{1}{2\pi} \int dt du \sqrt{1+\frac{(u/R)^{7-p}}
{1+u^2L_{\rm eff}^2}(\partial_ux)^2}\;,
\label{WILAC}
\ee
where $L_{\rm eff}^2 = n^T M^T M n$ is effective dipole vector.
First of all from this equation we find that despite the theory is
non-local, the end points of string can be
fixed at large $u$. We note that in the noncommutative gauge theory
where we have a non-zero B field with both legs along
the brane worldvolume we have problem for fixing the end points 
\cite{MR}, though we could fix it using moving frame \cite{AOJ}.

The action (\ref{WILAC}) is minimized when
\be 
\frac{(u/R)^{7-p}}{1+u^2L_{\rm eff}^2}\;\frac{\partial_u x}
{\left(1+\frac{(u/R)^{7-p}}
{1+u^2L_{\rm eff}^2}(\partial_ux)^2\right)^{1/2}} ={\rm constant}
= \left(\frac{(u_0/R)^{7-p}}
{1+u_0^2L_{\rm eff}^2}\right)^{1/2},
\ee
where $u_0$ is the point where $\partial_ux|_{u_0}\rightarrow \infty$. 
This equation can be solved for $\partial_ux$, that is
\be
\partial_ux=\frac{\left(\frac
{1+u^2L_{\rm eff}^2}{(u/R)^{7-p}}\right)^{1/2}}{\sqrt{\left(
{u\over u_0}\right)^{7-p}\frac{1+u^2_0L_{\rm eff}^2}{1+u^2L_{\rm eff}^2}-1
}}\;.
\ee
Hence
\be
x(u)=\int_{u_0}^u du\;\frac{\left(\frac
{1+u^2L_{\rm eff}^2}{(u/R)^{7-p}}\right)^{1/2}}{\sqrt{\left(
{u\over u_0}\right)^{7-p}\frac{1+u^2_0L_{\rm eff}^2}{1+u^2L_{\rm eff}^2}-1
}}\;.
\ee
The $Q \bar{Q}$ separation is defined by
\be
{l\over 2}:=x(u\rightarrow \infty)= \frac{R^{(7-p)/2}}
{u_0^{(5-p)/2}}\int_{1}^{\infty}dy
 \frac{(1+y^2{\bar L}_{\rm eff}^2)^{1/2}}{y^{(7-p)/2}\sqrt{
y^{7-p}\frac{1+{\bar L}_{\rm eff}^2}{1+y^2{\bar L}_{\rm eff}^2}-1}}\;.
\ee
Here ${\bar L}_{\rm eff}=u_0L_{\rm eff}$. Using (\ref{WILAC}) we 
can calculate the
energy of the $Q \bar{Q}$ system as following
\be
E={u_0\over \pi}[\int_1^{\infty}dy(
\frac{\left(y^{7-p}\frac{1+{\bar L}_{\rm eff}^2}{1+y^2{\bar L}_{\rm eff}^2}
\right)^{1/2}}
{\sqrt{y^{7-p}\frac{1+{\bar L}_{\rm eff}^2}{1+y^2{\bar L}_{\rm eff}^2}-1}}-1
)-1]\;.
\ee
Here we subtracted the infinity coming from mass of the 
W-boson which corresponds to string stretching all 
the way to $u=\infty$.

When the distance between quark and anti-quark is much bigger 
than their  dipole size, the energy is given by
\be
E\sim -\left(\frac{g_{\rm YM}^2 N}{l^2}\right)^{1/(5-p)}\left(1+
c_0L_{\rm eff}^2
\left(\frac{g_{\rm YM}^2 N}{l^2}\right)^{2/(5-p)}+\dots\right)\;,
\ee
where $c_0$ is a numerical constant.
The first term in the above expression is what we have 
in the ordinary gauge theory \cite{BISY} and the
second term can be interpreted as the dipole-dipole 
interaction\footnote{The Wilson loop of the dipole 
gauge theory has been also considered
by M.M. Sheikh-Jabbari \cite{JAB}.}.
 
\section{Conclusion} 
In this paper we have studied the D$_p$-branes supergravity solution in the
presence of non-zero B field with one one leg along the worldvolume
and the other transverse to it. We have shown that for $p\leq 5$ the
theory one the brane decouples form the gravity, therefore the 
supergravity solution provides a dual description of
worldvolume theory which is noncommutative dipole gauge theory.
This supergravity solution, in general, breaks the supersymmetry.
Nevertheless one can consider the case in which we left with some
supersymmetries. In this paper we only considered those brane solutions
which preserve 8 supercharges. Having a general gravity solution 
it is not clear if the solution is stable. Nevertheless since 
we are considering those brane systems which are supersymmetric
the solutions we found are stable.

We also studied the type II NS5-branes in the presence of
non-zero RR field with one leg along the transverse direction
to the NS5-branes and the others along their worldvolume.
These solutions can be thought as a deformation of little
string field theory. These theories also exhibit the same 
mass gap as little string theory. 

We have also studied the M5-brane in the presence of a
C field with two legs along the worldvolume and one leg
along the transverse directions. Up on compactification
on a circle we will find the 10-dimensional supergravity 
solution studied in this paper. We note, however, that
there is an other solution which we have not considered 
in this paper. This solution can be found by compactifying 
the M5-brane solution (\ref{M5D}) on a worldvolume direction 
in which  the C field is not defined. In the type IIA
limit this corresponds to a D4-brane solution in the presence 
of RR 3-form with two legs along the worldvolume and 
one leg transverse to it. There are also other
interesting solutions both in type II string theories
and M-theory where one leg of B field or C field is
along the time direction. We will back to these
questions in our incoming paper \cite{AGJY}.

We have also computed the Wilson loop of the dipole gauge theory
using  dual gravity description. We saw that despite the
fact that the theory is non-local we can fix the
end points of strings at large $u$. Note that
this is not the case for noncommutative field theory.
Therefore we might conclude that the dipole gauge theory
is a non-local deformation of ordinary gauge theory which
could be much more simpler than the non-local deformation
obtained by making the space to be
noncommutative, {\it i.e.} noncommutative gauge theory.

{\large {\bf Acknowledgments}}

We would like to thank H. Arfaei and M. R. Garousi for 
discussions and M. M. Sheikh-Jabbari for comments and discussions.

\end{document}